\documentclass[12pt,article,prc,aps,showpacs,showkeys,groupedaddress,floatfix]{revtex4}

\usepackage{epsfig,dcolumn,graphics,graphicx,amssymb,amsmath,bm}

\begin{document}

\baselineskip=14pt plus 1pt minus 1pt  

\title{Electric quadrupole transitions of the Bohr Hamiltonian with the Morse potential}

\author{I. Inci $^{1,3}$, D. Bonatsos $^2$, and I. Boztosun $^3$}
\affiliation{$^1$ I.N.F.N. and Dipartimento di Fisica ``Galileo Galilei", Universit\'a di Padova, I-35131, Padova, Italy}
\affiliation{$^2$ Institute of Nuclear Physics, National Centre for Scientific Research ``Demokritos'', GR-15310 Aghia Paraskevi, Attiki, Greece}
\affiliation{$^3$ Department of Physics, Akdeniz University, TR-07058, Antalya, Turkey}

\begin{abstract}

Eigenfunctions of the collective Bohr Hamiltonian with the Morse
potential have been obtained by using the Asymptotic Iteration Method
(AIM) for both $\gamma$-unstable and rotational structures.
$B(E2)$ transition rates have been calculated and compared to experimental data.
Overall good agreement is obtained for transitions within the ground state band, while
some interband transitions appear to be systematically underpredicted in $\gamma$-unstable nuclei
and overpredicted in rotational nuclei.

\end{abstract}

\keywords{Bohr Hamiltonian, Morse potential, electric quadrupole transitions}
\pacs{21.30.Fe, 21.60.Ev, 21.60.Fw, 21.10.Re}
\maketitle

\section{Introduction}

Shape phase transitions in atomic nuclei have attracted much attention
in the last decade \cite{RMP}, following the
introduction of the critical point symmetries E(5) \cite{iachelloPRL85} and
X(5) \cite{iachelloPRL87}. E(5) describes the second-order
phase transition point between vibrational and $\gamma$-unstable
nuclei, the parameter-free (up to overall scale factors) properties of such a
structure obtained by using an infinite-well potential in the
collective Bohr Hamiltonian \cite{bohr}. The first-order phase
transition point between vibrational and axially symmetric prolate
deformed rotational nuclei is given by X(5), the structural
properties again being found from a solution of the Bohr Hamiltonian with an
infinite-well potential \cite{iachelloPRL87}.

One way to describe nuclei which are close to or away from
these critical points is getting special solutions of the Bohr
Hamiltonian with a suitable potential. By using Coulomb-like and
Kratzer-like \cite{Kratzer} potentials in the Bohr Hamiltonian, eigenvalues and
eigenfunctions have been obtained in closed forms for the
$\gamma$-unstable region \cite{fortunatoJPG29}. In the deformed
rotational region,  analytical solutions of the Bohr Hamiltonian
using a Coulomb-like or Kratzer-like potential for the
$\beta$-part of the potential and a harmonic oscillator
potential for the $\gamma$-part of the potential have been found by
assuming that the Hamiltonian is separable into its variables
\cite{fortunatoJPG30}. For both regions, $\beta^{2n}$-type ($n=1,2,3,4$)
\cite{ariasPRC68,bonatsosPRC69,bonatsosPLB649} and
Davidson \cite{davidson,elliott,rowe,bonatsosPLB584,bonatsosPRC76}
potentials have been used in the Bohr Hamiltonian and
spectra and $B(E2)$ transition rates have been calculated.

The Morse potential has been used recently \cite{boztosunPRC77}
in order to solve the over-prediction of the energy
spacing problem within the $\beta$-band of X(5) and related
solutions \cite{castenPRL87,kruckenPRL88,castenJPG34}.
Closed expressions for the energy eigenvalues have been obtained
using the asymptotic iteration method (AIM) \cite{ciftciJPA36,ciftciJPA38}.
To complete this project, in the present work the eigenfunctions
are constructed and $B(E2)$ transition rates are calculated
and compared to the experimental data in the $\gamma$-unstable
and rotational regions, as well as to theoretical predictions
of other models.

The paper has the following structure. In Sec. \ref{aim}
the asymptotic iteration method (AIM) is briefly described. In Sec.
\ref{be2} the wave functions are constructed and the electric
quadrupole transition strengths are calculated in the $\gamma$-unstable and rotational
regions. Numerical results are given in Sec. \ref{results},
while in Sec. \ref{conclusions} the conclusions and outlook are discussed.
Details of the calculations are given in Appendix A.

\section{Overview of the Asymptotic Iteration Method}\label{aim}

The Asymptotic Iteration Method (AIM) has been proposed
\cite{ciftciJPA36,ciftciJPA38} and applied
\cite{karakoc,aygun,bayrakscripta,boztosunJMA47,soylu,bayrakJMS802,bayrakJPA39}
to the solution of second-order differential equations
of the form
\begin{equation}\label{differential}
  y''=\lambda_{0}(x)y'+s_{0}(x)y,
\end{equation}
where $\lambda_{0}(x)\neq 0$ and the prime denotes the derivative
with respect to $x$. The functions $s_{0}(x)$ and $\lambda_{0}(x)$
must be sufficiently differentiable. Eq. (\ref{differential}) has a general solution \cite{ciftciJPA36}
\begin{equation}\label{generalsolution}
  y(x)=\exp \left( - \int^{x} \alpha(x_{1}) dx_{1}\right ) \left [C_{2}+C_{1}
  \int^{x}\exp  \left( \int^{x_{1}} [\lambda_{0}(x_{2})+2\alpha(x_{2})] dx_{2} \right ) dx_{1} \right]
\end{equation}
for sufficiently large $k$, $k>0$, if
\begin{equation}\label{termination}
\frac{s_{k}(x)}{\lambda_{k}(x)}=\frac{s_{k-1}(x)}{\lambda_{k-1}(x)}=\alpha(x),
\end{equation}
where
\begin{eqnarray}\label{iteration}
  \lambda_{k}(x) & = &
  \lambda_{k-1}'(x)+s_{k-1}(x)+\lambda_{0}(x)\lambda_{k-1}(x), \quad
  \nonumber \\
s_{k}(x) & = & s_{k-1}'(x)+s_{0}(x)\lambda_{k-1}(x), \quad \quad
\quad \quad k=1,2,3,\ldots
\end{eqnarray}
For a given potential, the radial Schr\"{o}dinger equation is
converted to the form of Eq. (\ref{differential}). Then, s$_{0}(x)$
and $\lambda_{0}(x)$ are determined, and the functions s$_{k}(x)$ and
$\lambda_{k}(x)$ are calculated by the recurrence
relations of Eq. (\ref{iteration}).

The termination condition of the method, given in Eq. (\ref{termination}),
can be arranged as
\begin{equation}\label{quantization}
  \Delta_{k}(x)=\lambda_{k}(x)s_{k-1}(x)-\lambda_{k-1}(x)s_{k}(x)=0, \quad \quad
k=1,2,3,\ldots
\end{equation}
Then, the energy eigenvalues are obtained from the roots
of Eq. (\ref{quantization}) if the problem is exactly solvable. If not, for
a specific principal quantum number $n$, we choose a suitable $x_0$
point, generally determined as the maximum value of the asymptotic
wave function or the minimum value of the potential
\cite{ciftciJPA36,boztosunJMA47}, and the approximate energy eigenvalues
are obtained from the roots of this equation for sufficiently large
values of $k$ by iteration.

The corresponding eigenfunctions can be derived from the following
wave function generator for exactly solvable potentials
\begin{equation}\label{generator}
y_n (x) = C_2 \exp \left( { - \int\limits^x
{\frac{s_{n}(x_{1})}{\lambda_{n}(x_{1})}dx_{1}} } \right),
\end{equation}
where $n$ represents the radial quantum number.

Recently, Boztosun and Karakoc \cite{boztosunCPL24} have further improved the method for the exactly solvable problems.
This improved form had been applied to the exactly separable $\gamma\approx 0$ solution of the Morse potential in Ref. \cite{boztosunPRC77}.

\section{ B(E2) transition strengths}\label{be2}

\subsection{The $\gamma$-unstable case}

The Bohr Hamiltonian \cite{bohr} is
\begin{equation}
H=-\frac{\hbar^2}{2B}\left[\frac{1}{\beta^4}\frac{\partial}{\partial\beta}\beta^4\frac{\partial}{\partial\beta}+\frac{1}{\beta^2\sin3\gamma}\frac{\partial}{\partial\gamma}\sin3\gamma\frac{\partial}{\partial\gamma}
-\frac{1}{4\beta^2}\sum_{k=1}^3\frac{\hat{Q}_k^2}{\sin^2(\gamma-\frac{2\pi}{3}k)}\right]+V(\beta,\gamma),
\end{equation}
where $\beta$ and $\gamma$ are collective coordinates describing the deformation and the shape of a nucleus, while $\hat{Q}_k$ represents the angular momentum component in the body-fixed coordinate system and $B$ is the mass parameter. For $\gamma$-unstable structure, the potential energy is independent of $\gamma$, namely $V(\beta, \gamma)=V(\beta)$, and by assuming a wave function of the form \cite{wilets}
\begin{equation}\label{unstable-wf}
\psi(\beta, \gamma, \theta_i)=\xi(\beta)\Phi(\gamma, \theta_i),
\end{equation}
where $\theta_i$ ($i=1$, 2, 3) are the Euler angles,
one can achieve separation of variables
\begin{eqnarray}
\nonumber
\left[-\frac{1}{\sin3\gamma}\frac{\partial}{\partial\gamma}\sin3\gamma\frac{\partial}{\partial\gamma}
+\frac{1}{4}\sum_{k=1}^3\frac{\hat{Q}_k^2}{\sin^2(\gamma-\frac{2\pi}{3}k)}\right]\Phi(\gamma, \theta_i) & = & \tau(\tau+3)\Phi(\gamma, \theta_i),\\
\left[-\frac{1}{\beta^4}\frac{\partial}{\partial\beta}\beta^4
\frac{\partial}{\partial\beta}+\frac{\tau(\tau+3)}{\beta^2}+u(\beta)\right]\xi(\beta)& = & \epsilon \xi(\beta),
\end{eqnarray}
where $\epsilon={2B\over\hbar^2}E$ and $u(\beta)={2B\over\hbar^2}V(\beta)$ are the reduced energies and potentials, respectively,
while $\tau$ is the seniority quantum number \cite{Rakavy}.

In the radial part of the Hamiltonian for $u(\beta)$ the Morse potential \cite{morse} is used
\begin{equation}
u(\beta)=e^{-2a(\beta-\beta_e)}-2e^{-a(\beta-\beta_e)},
\end{equation}
leading to the energy spectrum \cite{boztosunPRC77}
\begin{equation}
\epsilon_{n,\tau}={\nu c_0\over\beta_e^2}-\left[{\gamma_1^2\over2\beta_e\gamma_2}-\left(n+{1\over2}\right){\alpha\over\beta_e}\right]^2,
\end{equation}
where $n$ is the principal quantum number, and
\begin{eqnarray}\label{CCs}
c_0=1-{3\over\alpha}+{3\over\alpha^2},\quad
c_1={4\over\alpha}-{6\over\alpha^2},\quad
c_2=-{1\over\alpha}+{3\over\alpha^2}, \nonumber \\
 \alpha = a \beta_e,\quad
\gamma_1^2= 2\beta_e^2-\nu c_1,\quad \gamma_2^2= \beta_e^2+\nu c_2,
\quad \nu= \tau(\tau+3)+2.
\end{eqnarray}

To get the radial wave function one needs the parametrization \cite{boztosunPRC77}
\begin{eqnarray}\label{unstablemorse-wf}
\xi(\beta)=\beta^{-2}\chi(\beta), \quad x={\beta-\beta_e\over\beta_e}, \quad y=e^{-\alpha x}, \quad \chi_{n, \tau}(y)= y^{K_{n, \tau}/\alpha} e^{-\gamma_2 y/\alpha} f_{n, \tau}(y), \nonumber \\
K_{n,\tau}=\sqrt{\nu c_0- \epsilon_{n,\tau} \beta_e^2},
\end{eqnarray}
leading to \cite{boztosunPRC77}
\begin{equation}\label{unstable-differential}
f_{n, \tau}^{''}(y)=\left({2\gamma_2\alpha y-2\alpha K_{n,\tau}-\alpha^2\over\alpha^2y}\right)f_{n, \tau}^{'}(y)+ \left({2K_{n,\tau}\gamma_2+\alpha\gamma_2-\gamma_1^2\over\alpha^2y}\right) f_{n, \tau}(y).
\end{equation}
As described in Appendix A, the solution of this equation is found to be
\begin{equation}
f_{n, \tau}(y)=N_{n, \tau}L_n^{2K_{n, \tau}/\alpha}\left({2\gamma_2\over\alpha}y\right),
\end{equation}
where $L$ represents the Laguerre polynomials and $N$ are normalization constants determined from the normalization condition,
\begin{equation}
\int_0^{\infty} \xi^2(\beta) \beta^4 d\beta = \int_0^{\infty} \chi^2(\beta) d\beta={1\over a}\int_0^{e^{\alpha}} {1\over y}\chi^2(y) dy = 1,
\end{equation}
leading to
\begin{equation}
N_{n, \tau}= \left[{a(\omega-2n-1)n!\over(\omega-n-1)\Gamma(\omega-n-1)}\left({2\gamma_2\over\alpha}\right)^{\omega-2n-1}\right]^{1/2},
\quad \omega ={\gamma_1^2 \over \alpha\gamma_2}.
\end{equation}

The quadrupole operator is defined as \cite{wilets}
\begin{equation}\label{TE2}
T_\mu^{(E2)}=t\alpha_\mu=t\beta\left[D_{\mu, 0}^{(2)}(\theta_i)\cos \gamma+{1\over\sqrt2}\left(D_{\mu, 2}^{(2)}(\theta_i)+D_{\mu, -2}^{(2)}(\theta_i)\sin \gamma\right)\right],
\end{equation}
where $t$ is a scale factor, while $D(\theta_i)$ represent the Wigner functions of the Euler angles $\theta_i$.
Then $B(E2)$ transition rates are given by
\begin{eqnarray}
\nonumber
B(E2; s_i,L_i \rightarrow s_f,L_f) & = & {1\over 2L_i+1}|\langle s_f, L_f\|T^{(E2)}\|s_i,L_i\rangle|^2 \\
& = & {2L_f+1\over 2L_i+1}B(E2; s_f,L_f \rightarrow s_i,L_i).
\end{eqnarray}
As mentioned in Ref. \cite{bonatsosPRC69}, the states defined in Eq. (\ref{unstable-wf}) with $\nu_{\Delta}=0$
(where $\nu_{\Delta}$ is the missing quantum number in the SO(5)$\supset$SO(3) reduction) and $L=2\tau$ take the form
\begin{equation}
\psi(\beta, \gamma, \theta_i)=\xi(\beta)\phi_\tau(\gamma, \theta_i),
\end{equation}
where the functions $\phi_\tau(\gamma, \theta_i)$ are given by \cite{bes}
\begin{equation}
\phi_\tau(\gamma, \theta_i)= {1 \over 4\pi} \sqrt{(2\tau+3)!!\over  \tau!}\left(\alpha_2\over\beta\right)^\tau,
\end{equation}
with $\alpha_2$ defined in Eq. (\ref{TE2}).
Then $B(E2)$s are found to be \cite{bonatsosPRC69}
\begin{eqnarray}
B\left(E2; L_{n, \tau} \rightarrow (L+2)_{n^{'}, \tau+1}\right) & = & {(\tau+1)(4\tau+5)\over(2\tau+5)(4\tau+1)}t^2 I^2_{n^{'},\tau+1;n,\tau},
\quad L=2\tau,  \\
B\left(E2; (L+2)_{n^{'}, \tau+1} \rightarrow  L_{n, \tau}\right) & = & {(\tau+1)\over(2\tau+5)}t^2 I^2_{n^{'},\tau+1;n,\tau},   \quad L=2\tau,
\end{eqnarray}
where
\begin{equation}
I_{n^{'},\tau+1;n,\tau}=\int_0^\infty \beta \chi_{n^{'},\tau+1}(\beta)\chi_{n,\tau}(\beta)d\beta .
\end{equation}
For the Morse potential, using the eigenfunctions in Eq.(\ref{unstablemorse-wf}) one gets
\begin{eqnarray}
\nonumber
I_{n^{'},\tau+1;n,\tau} & = & \int_0^\infty \beta^5 \xi_{n^{'},\tau+1}(\beta)\xi_{n,\tau}(\beta)d\beta \\
& = & {1\over a^2} \int_0^{e^\alpha}\left(\alpha-\ln y\over y\right)\chi_{n^{'},\tau+1}(y)\chi_{n,\tau}(y)dy.
\end{eqnarray}


\subsection{The rotational case}

For the rotational region, the reduced potential is $\gamma$-dependent. Exact separation of the Bohr Hamiltonian can be achieved
\cite{wilets} by taking the reduced potential to be of the form of $u(\beta,\gamma)=u(\beta)+{v(\gamma)/\beta^2}$. Then
the wave functions can be written as
\begin{equation}
\psi(\beta,\gamma,\theta_i)=\xi_L(\beta)\Gamma_K(\gamma)D_{M,K}^L(\theta_i),
\end{equation}
where $L$ is the angular momentum quantum number, while $M$ and $K$
are the angular momentum projections on the laboratory-fixed $z$-axis and body-fixed $z^{'}$-axis, respectively.
For an axially-symmetric prolate deformed structure,
assuming $v(\gamma)=(3c)^2\gamma^2$,  the $\gamma$-part of the solution \cite{bonatsosPRC76} is given in terms of Laguerre polynomials
\begin{equation}\label{Gammawf}
\Gamma_{n_\gamma,K}(\gamma)=N_{n_\gamma,K}\gamma^{|K/2|}e^{-3c\gamma^2/2}L_{\tilde{n}}^{|K/2|}(3c\gamma^2)
\end{equation}
with
\begin{equation}\label{Epsilongamma}
 \quad n_\gamma=0,1,2..., \qquad \tilde{n}={n_\gamma-|K/2|\over2},
\end{equation}
while the normalization constants $N_{n_\gamma,K}$ can be found from the normalization condition
\begin{equation}\label{gammanorm}
\int_0^{\pi/3}\Gamma^2_{n_\gamma,K}(\gamma)|\sin 3\gamma| d\gamma=1.
\end{equation}
For deformed nuclei one has $\gamma\approx0$, which implies that one can use $|\sin 3\gamma|\simeq3|\gamma|$. The normalization constant
for the $(n_{\gamma},K)=(0,0)$ state is found to be
\begin{equation}\label{N00}
N_{0,0}=\sqrt{\frac{2c}{1-e^{-\frac{c\pi^2}{3}}}}.
\end{equation}
For high $c$ (say $c\geq2$) the $e^{-\frac{c\pi^2}{3}}$ term can be neglected, since it becomes of the order of $10^{-4}$, or smaller. Then the normalization constants are as follows,
\begin{eqnarray}\label{gammanorconst}
\nonumber
  N_{0,0} = \sqrt{2c}        & \quad , \quad & N_{3,6} = \sqrt{9c^4},\\\nonumber
  N_{1,2} = \sqrt{6c^2}      & \quad , \quad & N_{4,0} = \sqrt{2c}, \\\nonumber
  N_{2,0} = \sqrt{2c}        & \quad , \quad & N_{4,4} = \sqrt{3c^3}, \\\nonumber
  N_{2,4} = \sqrt{9c^3}      & \quad , \quad & N_{4,8} = \sqrt{\frac{27}{4}c^5}, \\
  N_{3,2} = \sqrt{3c^2}      & \quad , \quad & N_{5,10} = \sqrt{\frac{81}{20}c^6}.
\end{eqnarray}
The energy spectrum turns out to be \cite{boztosunPRC77}
\begin{equation}
\epsilon_{n,L}={\mu
c_0\over\beta_e^2}-\left[{\gamma_1^2\over2\beta_e\gamma_2}-\left(n+{1\over2}\right){\alpha\over\beta_e}\right]^2,
\end{equation}
where
\begin{eqnarray}
\gamma_1^2= 2\beta_e^2-\mu c_1,\quad \gamma_2^2= \beta_e^2+\mu c_2, \nonumber\\
\mu={L(L+1)\over3}+2+\lambda, \quad  \lambda = \epsilon_\gamma -{K^2 \over 3},
 \quad \epsilon_\gamma = (3C) (n_\gamma+1), \quad C=2c,
 \end{eqnarray}
while the rest of the symbols have the same value as in Eq. (\ref{CCs}).

To get the radial wave function, one needs the parametrization
\begin{eqnarray}\label{R-wavefunction}
\xi_{n,L}(\beta)=\beta^{-2}\chi_{n,L}(\beta), \quad x={\beta-\beta_e\over\beta_e}, \quad y=e^{-\alpha x},
\quad \chi_{n,L}(y)=y^{\rho_{n,L}/\alpha}e^{-\gamma_2 y /\alpha }R_{n,L}(y), \nonumber \\
\rho_{n,L}={\gamma_1^2\over2\gamma_2}-{(n+{1\over2})\alpha},
 \end{eqnarray}
leading to
\begin{equation}\label{R-function}
R_{n,L}^{''}(y)=\left({2\gamma_2\alpha y-2\alpha \rho_{n,L}-\alpha^2\over\alpha^2y}\right)R_{n, L}^{'}(y)+ \left({2\rho_{n,L}\gamma_2+\alpha\gamma_2-\gamma_1^2\over\alpha^2y}\right) R_{n, L}(y).
\end{equation}
In the right-hand-side of Eq. (\ref{R-function}), the first and the second functions in parentheses
are $\lambda_0(y)$ and $s_0(y)$ of Eq. (\ref{differential}).
After determination of $\lambda_k$'s and $s_k$'s, through the use of Eq. (\ref{iteration}),
the wave functions are found by using Eq. (\ref{generator}) to be
\begin{equation}
R_{n, L}(y)=N_{n,L}L_n^{2\rho_{n,L}/\alpha}\left({2\gamma_2\over\alpha}y\right),
\end{equation}
where $L$ represents the Laguerre polynomials, while $N_{n,L}$ correspond to normalization constants,
obtained from the normalization condition
\begin{equation}
\int_0^{\infty} \xi^2(\beta) \beta^4 d\beta = 1
\end{equation}
to be
\begin{equation}
N_{n, L}= \left[{a(\sigma-2n-1)n!\over(\sigma-n-1)\Gamma(\sigma-n-1)}\left({2\gamma_2\over\alpha}\right)^{\sigma-2n-1}\right]^{1/2}, \qquad
\sigma ={\gamma_1^2 \over\alpha\gamma_2}.
\end{equation}
Then one can calculate the B(E2) transition rates
by following the procedure given in Appendix B of Ref. \cite{bonatsosPRC76}
\begin{equation}
B(E2;nLn_\gamma K\rightarrow n^{'}L^{'}n^{'}_\gamma K^{'})={5\over16\pi}t^2\left(\langle L, 2, L^{'}|K, K^{'}-K,K^{'}\rangle\right)^2 B^2_{n,L,n^{'},L^{'}}C^2_{n_\gamma,K,n^{'}_\gamma, K^{'}}
\end{equation}
In our case,  $B_{n,L,n^{'},L^{'}}$  is given by
\begin{equation}
B_{n,L,n^{'},L^{'}}=-{1\over a^2}\int_{e^{\alpha}}^0{\alpha-ln(y)\over y}\chi_{n,L}(y)\chi_{n^{'},L^{'}}(y) dy,
\end{equation}
where $\chi_{n,L}(y)$ is defined in Eq. (\ref{R-wavefunction}). The other constant $C_{n_\gamma,K,n^{'}_\gamma, K^{'}}$ is
\begin{eqnarray}
\nonumber
C_{n_\gamma,K,n^{'}_\gamma, K^{'}} & = & \int_0^{\pi/3}\cos\gamma \Gamma_{n^{'}_\gamma,K^{'}}(\gamma)\Gamma_{n_\gamma,K}(\gamma)|\sin 3\gamma| d\gamma,  \quad \Delta K=0,  \\
& = & \int_0^{\pi/3}\sin\gamma \Gamma_{n^{'}_\gamma,K^{'}}(\gamma)\Gamma_{n_\gamma,K}(\gamma)|\sin 3\gamma| d\gamma,  \quad\Delta K=2.
\end{eqnarray}
These integrals can be calculated by remembering that for deformed nuclei $\gamma\approx0$,  implying $\cos \gamma\simeq1$ and $\sin \gamma\simeq\gamma$. Under these approximations, for $\Delta K=0$ transitions one has $C_{n_\gamma,K,n^{'}_\gamma, K^{'}}=\delta_{n_\gamma,n^{'}_\gamma}\delta_{K,K^{'}}$,
therefore
\begin{equation}
B\left(E2;nLn_{\gamma}K\rightarrow n^{'}L^{'}n_{\gamma}K\right) \approx
{5\over 16\pi}t^2\left(<L2L^{'}|K0K>\right)^2 B^2_{n,L,n^{'},L^{'}}.
\end{equation}

For $\Delta K=2$ transitions, if the states have $\tilde{n}=0$ [see Eq. (\ref{Epsilongamma})], then the Laguerre polynomials are unity
[see Eq. (\ref{Gammawf})] and
\begin{equation}
C_{n_\gamma,K,n^{'}_\gamma, K^{'}}=3N_{n_\gamma^{'},K^{'}}N_{n_\gamma,K}\int_0^{\pi/3}\gamma^{\frac{4+|K^{'}|+|K|}{2}} e^{-3c\gamma^2} d\gamma.
\end{equation}
As an example, we give $C_{n_\gamma,K,n^{'}_\gamma, K^{'}}$ for the $(n^{'}_\gamma=1, K^{'}=2)\rightarrow(n_\gamma=0,K=0)$ transition,
\begin{eqnarray}
\nonumber
C_{0,0,1,2}& = & 3N_{0,0}N_{1,2}\int_0^{\pi/3}\gamma^{3} e^{-3c\gamma^2} d\gamma\\
& = & 3N_{0,0}N_{1,2}\left(\frac{1}{18c^2}-\left[\frac{1}{18c^2}+\frac{\pi^2}{54c}\right]e^{-\frac{c\pi^2}{3}}\right).
\end{eqnarray}
As explained below Eq. (\ref{N00}), the second term can be neglected. Then
\begin{equation}
C_{0,0,1,2}= \frac{1}{6c^2}N_{0,0}N_{1,2}.
\end{equation}
The $\gamma$-integrals for higher $\Delta K=2$ transitions are
\begin{eqnarray}
\nonumber
C_{1,2,2,4} & = & \frac{1}{9c^3}N_{1,2}N_{2,4},\\\nonumber
C_{2,4,3,6} & = & \frac{1}{9c^4}N_{2,4}N_{3,6},\\\nonumber
C_{3,6,4,8} & = & \frac{4}{27c^5}N_{3,6}N_{4,8},\\
C_{4,8,5,10} & = & \frac{20}{81c^6}N_{4,8}N_{5,10}.
\end{eqnarray}
In the above, the normalization factors $N_{n_\gamma,K}$ are given in
Eq. (\ref{gammanorconst}).

\section{Numerical results}\label{results}

Numerical calculations for $B(E2)$s have been performed for 30 $\gamma$-unstable nuclei,
listed in the left part of Table \ref{MorsePar}, as well as for 32 rotational nuclei,
listed in the right part of Table \ref{MorsePar}. The parameters listed for these nuclei
have been obtained in Ref. \cite{boztosunPRC77} by fitting the experimental energy spectra.
No attempt to include $B(E2)$s in the fit has been made. Nuclei for which  at least
one $B(E2)$ value is known in addition to $B(E2: 2_1^+\to 0_1^+)$
(to which we normalize the other $B(E2)$ values) are included in the tables.

$B(E2)$s for $\gamma$-unstable nuclei are shown in Table \ref{gammabe2},
while $B(E2)$s for rotational nuclei are shown in Table \ref{rotationalbe2}.
It is clear from Table \ref{gammabe2} that the Morse potential
gives in general good results for the transitions within the ground state band
of $\gamma$-unstable nuclei, while the predictions for some interband transitions
are systematically lower than the experimental data.
On the other hand, in Table \ref{rotationalbe2} one can see that
the Morse potential is successful in predicting the intraband $B(E2)$s
within the ground state band of deformed nuclei, while most inter-band $B(E2)$s are
systematically overpredicted.

Numerical results for the spectra and $B(E2)$s of the same nuclei,
both in the $\gamma$-unstable and in the rotational regions,
have been provided recently \cite{PLB683,DDMDav} using a Bohr Hamiltonian in which the
mass is not a constant, as in the present work, but it is a function
of the deformation, with a Davidson potential used in the radial
degree of freedom. Analytical solutions are obtained through
the use of techniques of supersymmetric quantum mechanics (SUSYQM) \cite{SUSYQM}.
It is instructive to compare the results of the present approach
and the deformation-dependent mass (DDM) Davidson approach.
The following remarks apply.

1) In both solutions  the same assumptions
have been made as far as the separability of the potential is concerned.

2) In both solutions the same number of free parameters (up to overall scale factors)
is used both in the $\gamma$-unstable case (two parameters) and in the rotational case
(three parameters).

3) The same set of experimental data has been used in both models.

4) In both models the parameters have been determined by fitting the energy levels alone.
These parameters have been subsequently used for calculating the $B(E2)$s.

5) In both models the problem of large spacings within the $\beta$-band is well accounted for.

6) In both models both intraband and interband $B(E2)$s for $\gamma$-unstable nuclei
are somewhat systematically overpredicted. The intraband $B(E2)$s for rotational nuclei
predicted by both models are in general in good agreement to the data, while some interband
$B(E2)$s are still systematically overpredicted.

The similarity of the numerical results provided by the two models invites for a look into
their differences.

1) In the DDM Davidson framework \cite{DDMDav} it seems that the dependence of the mass on the deformation
is the main factor leading to improved agreement to experiment, since the increase of the moments of inertia
with deformation is moderated, as one can see by comparing the results of Ref. \cite{DDMDav} to these of Ref. \cite{bonatsosPRC76},
where the Davidson potential with constant mass is considered. The mass dependence on the deformation
is absent in the present Morse model.

2) In the present Morse model, it seems that it is the shape of the potential which leads to good agreement to experiment.
It is known that the rapid increase of the potential ``on the right hand side'', i.e., for large $\beta$, leads
to large spacings within the $\beta$ band, which are moderated if the increase of the potential is made slower \cite{sloped}.
The Morse potential becomes flat on the right hand side, while the Davidson potential is growing as $\beta^2$.

From the above it is plausible that a DDM form of the Kratzer potential \cite{Kratzer} should be studied, since this
will have both good features of the models already considered, namely the deformation dependence of the mass,
used in the Davidson model, and the finiteness at large $\beta$, possessed by the Morse model.
Work in this direction is in progress. It should be noticed that among the potentials soluble by SUSYQM
techniques \cite{SUSYQM} with deformation dependent mass \cite{Bagchi}, only the Davidson and Kratzer potentials
(i.e., harmonic oscillator and Coulomb potentials with centrifugal terms) can be successfully treated
in more than one dimensions, allowing the presence of angular momentum. As of today, the Morse potential can be treated
successfully by these techniques only in one dimension.

It should be mentioned that numerical solutions of the Bohr Hamiltonian avoiding the
assumptions guaranteeing the separation of variables used in the present approach,
can be obtained in the framework of the algebraic collective model \cite{Rowe,Turner,ACM}. Preliminary results \cite{CaprioPLB672}
indicate that the overprediction of interband $B(E2)$s seems to persist even when
separation of variables is avoided.

\section{Conclusions}\label{conclusions}

Using the asymptotic iteration method (AIM), the Bohr Hamiltonian with the Morse potential
has been studied for $\gamma$-unstable nuclei and for deformed nuclei. Wave functions have been obtained
in closed form and $B(E2)$ transition rates have been evaluated for 30 $\gamma$-unstable and 32 rotational
nuclei and compared to experimental data. Agreement to experimental data, as well as to theoretical
predictions obtained with a Bohr Hamiltonian possessing a deformation-dependent mass (DDM) and using a
Davidson potential is in general good for intraband transitions within the ground state band.
In the case of $\gamma$-unstable nuclei, some interband transitions are systematically underpredicted
by both models, while in rotational nuclei some interband $B(E2)$s are systematically overpredicted by both models.
In order to test if this systematic overprediction is due to the shape of the potential used
or to the special forms of the potentials used, which are amenable to separation of variables,
a DDM calculation using a Kratzer potential is suggested.

\section{acknowledgment}

This work has been partly supported by the Turkish Science and
Research Council (T\"{U}B\.{I}TAK) through Grant Numbers 109T373 and 110T388, the
Turkish Academy of Sciences (T\"{U}BA-GEB\.{I}P) and the Akdeniz
University Scientific Research Projects Unit.


\appendix
\section{$\gamma$-unstable case}

To get the solution of the second-order differential equation of the
form of Eq. (\ref{differential}), our starting point is Eq. (\ref{generator}).

Comparing Eq. (\ref{unstable-differential})
to Eq.(\ref{differential}),
  one can see that
\begin{eqnarray}
\nonumber
\lambda_0(y) & = & {2\gamma_2\alpha y-2\alpha K_{0,\tau}-\alpha^2\over\alpha^2y}, \\
 s_0(y) & = & {2K_{0,\tau}\gamma_2+\alpha\gamma_2-\gamma_1^2\over\alpha^2y},
\end{eqnarray}
where $K_{0,\tau}=(\gamma_1^2-\alpha\gamma_2)/2\gamma_2$ and
\begin{equation}
f_{0,\tau}(y)=(C_2)_{0,\tau}
\exp\left({-\int^y{s_0\over\lambda_0}dy^{'}}\right)=(C_2)_{0,\tau}.
\end{equation}
Then one can calculate $f_{n_,\tau}$ using the definition of
$\lambda_k$'s and $s_k$'s as given in Eq. (\ref{iteration}), finding
\begin{eqnarray}
\nonumber
\lambda_1 & = & \lambda_0^{'}(y)+s_0(y)+\lambda_0(y) \lambda_0(y), \quad  s_1 = s_0^{'}(y)+s_0(y)+\lambda_0(y), \quad K_{1,\tau}={\gamma_1^2-3\alpha\gamma_2\over2\gamma_2}, \\
f_{1,\tau}(y) & = & (C_2)_{1,\tau}
\exp\left({-\int^y{s_1\over\lambda_1}dy^{'}}\right)=(C_2)_{1,\tau}\left(2\gamma_2^2y+2\alpha\gamma_2-\gamma_1^2\right),
\end{eqnarray}
\begin{eqnarray}
\nonumber \lambda_2 & = & \lambda_1^{'}(y)+s_1(y)+\lambda_0(y)
\lambda_1(y), \quad  s_2 = s_1^{'}(y)+s_0(y)+\lambda_1(y),  \quad
K_{2,\tau}={\gamma_1^2-5\alpha\gamma_2\over2\gamma_2}, \\\nonumber
f_{2,\tau}(y) & = & (C_2)_{2,\tau} \exp\left({-\int^y{s_2\over\lambda_2}dy^{'}}\right)\\
& = & (C_2)_{2,\tau}\left(4\gamma_2^4 y^2+12\gamma_2^3\alpha
y-4\gamma_1^2
y\gamma_2^2+12\alpha^2\gamma_2^2-7\gamma_1^2\gamma_2\alpha+\gamma_1^4\right),
\end{eqnarray}
\begin{eqnarray}
\nonumber \lambda_3 & = & \lambda_2^{'}(y)+s_2(y)+\lambda_0(y)
\lambda_2(y), \quad  s_3 = s_2^{'}(y)+s_0(y)+\lambda_2(y),  \quad
K_{3,\tau}={\gamma_1^2-7\alpha\gamma_2\over2\gamma_2}, \\\nonumber
f_{3,\tau}(y) & = & (C_2)_{3,\tau} \exp\left({-\int^y{s_3\over\lambda_3}dy^{'}}\right) \\
& \quad & \ldots \quad {\rm etc}.
\end{eqnarray}  From
these expressions, it is easy to write $f_{n,\tau}(y)$ in a
closed form as
\begin{equation}
f_{n, \tau}(y)=(C_2)_{n, \tau}L_n^{2K_{n,
\tau}/\alpha}\left({2\gamma_2\over\alpha}y\right)
\end{equation}
where $L$ represents the Laguerre polynomials.

\newpage

\begin{table}

\caption{Morse potential parameters for nuclei in the  $\gamma$-unstable region (two columns on the left)
and for nuclei in the rotational region (two columns on the right),
taken from Ref.
\cite{boztosunPRC77}.}\label{MorsePar}
\bigskip

\begin{tabular}{c| c c || c| c c ||| c| c c c || c| c c c}

\hline
  nucl. & $\beta_e$ & $a$ & nucl. & $\beta_e$ & $a$ & nucl. & $\beta_e$ & $a$ & $C$ & nucl. & $\beta_e$ & $a$ & $C$\\
\hline
  $^{98} $Ru & 3.84 & 0.44 &$^{120}$Xe & 7.05 & 0.16&$^{154}$Sm & 8.4 & 0.20 &13.7 &$^{174}$Yb &17.2 & 0.05 & 15.1 \\
  $^{100}$Ru & 4.43 & 0.36 &$^{124}$Xe & 6.88 & 0.17&$^{156}$Gd & 7.5 & 0.22 &10.5 &$^{174}$Hf & 6.1 & 0.26 & 10.4 \\
  $^{102}$Ru & 3.78 & 0.42 &$^{128}$Xe & 4.74 & 0.39&$^{158}$Gd & 8.3 & 0.36 &10.6 &$^{176}$Hf & 7.9 & 0.18 & 11.7 \\
  $^{104}$Ru & 7.57 & 0.10 &$^{130}$Ba & 5.58 & 0.77&$^{158}$Dy & 6.7 & 0.25 & 7.1 &$^{178}$Hf & 8.5 & 0.16 &  9.2 \\
  $^{102}$Pd & 4.30 & 0.34 &$^{132}$Ba & 4.63 & 0.29&$^{160}$Dy & 9.2 & 0.18 & 8.2 &$^{182}$W  & 8.3 & 0.13 &  8.6 \\
  $^{104}$Pd & 4.15 & 0.41 &$^{134}$Ba & 3.82 & 0.50&$^{162}$Dy & 8.3 & 0.44 & 7.2 &$^{184}$W  & 6.6 & 0.15 &  5.5 \\
  $^{106}$Pd & 3.93 & 0.43 &$^{142}$Ba & 5.45 & 0.60&$^{164}$Dy &13.1 & 0.14 & 6.8 &$^{186}$W  & 5.3 & 0.22 &  4.2 \\
  $^{108}$Pd & 4.36 & 0.30 &$^{148}$Nd & 6.40 & 0.14&$^{162}$Er & 7.1 & 0.26 & 7.0 &$^{186}$Os & 5.6 & 0.27 &  4.2 \\
  $^{108}$Cd & 3.97 & 0.43 &$^{152}$Gd & 3.93 & 0.40&$^{164}$Er & 9.3 & 0.14 & 6.6 &$^{188}$Os & 5.3 & 0.34 &  2.8 \\
  $^{110}$Cd & 3.66 & 0.47 &$^{154}$Dy & 4.22 & 0.38&$^{166}$Er & 9.7 & 0.23 & 6.5 &$^{230}$Th & 7.6 & 0.19 & 10.7 \\
  $^{112}$Cd & 3.55 & 0.50 &$^{156}$Er & 4.75 & 0.34&$^{168}$Er & 8.1 & 0.59 & 6.7 &$^{232}$Th & 9.6 & 0.15 & 11.8 \\
  $^{114}$Cd & 3.43 & 0.51 &$^{192}$Pt & 6.42 & 0.19&$^{170}$Er & 6.5 & 0.17 & 9.0 &$^{234}$U  &12.1 & 0.12 & 15.5 \\
  $^{116}$Cd & 4.10 & 0.47 &$^{194}$Pt & 7.28 & 0.14&$^{166}$Yb & 6.8 & 0.23 & 6.7 &$^{236}$U  &13.8 & 0.10 & 15.0 \\
  $^{118}$Cd & 4.11 & 0.47 &$^{196}$Pt & 6.26 & 0.15&$^{168}$Yb & 7.9 & 0.19 & 8.0 &$^{238}$U  &13.9 & 0.10 & 17.6 \\
  $^{118}$Xe & 5.40 & 0.19 &$^{198}$Pt & 3.87 & 0.39&$^{170}$Yb & 7.8 & 0.49 & 9.2 &$^{238}$Pu & 9.8 & 0.50 & 15.9 \\
             &      &      &           &      &     &$^{172}$Yb & 7.8 & 0.17 &13.7 &$^{250}$Cf &12.8 & 0.20 & 16.7 \\
\hline
\end{tabular}
\end{table}

\newpage

\begin{table}

\caption{Comparison of experimental data \cite{NDS} (upper line) for several $B(E2)$ ratios of $\gamma$-unstable nuclei
to predictions (lower line) by the Bohr Hamiltonian with the Morse potential, for the
parameter values shown in Table \ref{MorsePar}.}\label{gammabe2}

\bigskip

\begin{tabular}{l r@{.}l r@{.}l r@{.}l r@{.}l r@{.}l r@{.}l r@{.}l r@{.}l r@{.}l r@{.}l}
\hline
   \multicolumn{1}{l}{nucl.}
   &\multicolumn{2}{c} {$4_1\to 2_1 \over 2_1\to 0_1$}
    &\multicolumn{2}{c} {$6_1\to 4_1 \over 2_1\to 0_1$}
    &\multicolumn{2}{c} {$8_1\to 6_1 \over 2_1\to 0_1$}
   &\multicolumn{2}{c} {$10_1\to 8_1 \over 2_1\to 0_1$}
    &\multicolumn{2}{c} {$2_2 \to 2_1 \over 2_1\to 0_1$}
   &\multicolumn{2}{c}{$2_2 \to 0_1 \over 2_1\to 0_1$}
   &\multicolumn{2}{c}{$0_2 \to 2_1 \over 2_1\to 0_1$}
   &\multicolumn{2}{c}{$2_3 \to 0_1 \over 2_1 \to 0_1$}  \\

   & \omit\span & \omit\span & \omit\span & \omit\span &
  \omit\span &  \multicolumn{2}{c} {x $10^3$} &  \omit\span &
  \multicolumn{2}{c} {x $10^3$}
   \\
\hline
$^{98}$Ru   & 1&44(25) & \omit\span & \omit\span & \omit\span &
1&62(61) &  36&0(152)      & \omit\span & \omit\span \\
           & 1&71 & 2&52 & 3&61 & 5&27 & 1&71 & 0&0 & 0&82 & 4&24  \\

$^{100}$Ru   & 1&45(13) & \omit\span & \omit\span & \omit\span &
0&64(12) &  41&1(52)      & 0&98(15) & \omit\span \\
           & 1&65 & 2&31 & 3&07 & 4&00 & 1&65 & 0&0 & 0&72 & 8&20 \\

$^{102}$Ru   & 1&50(24) & \omit\span & \omit\span & \omit\span &
0&62(7) &  24&8(7)      & 0&80(14) & \omit\span \\
           & 1&73 & 2&53 & 3&54 & 4&86 & 1&73 & 0&0 & 0&88 & 4&61 \\

$^{104}$Ru   & 1&18(28) & \omit\span & \omit\span & \omit\span &
0&63(15) &  35&0(84)      & 0&42(7) & \omit\span \\
           & 1&62 & 2&10 & 2&49 & 2&80 & 1&62 & 0&0 & 0&71 & 16&34 \\

$^{102}$Pd   & 1&56(19) & \omit\span & \omit\span & \omit\span &
0&46(9) &  128&8(735)      & \omit\span & \omit\span \\
           & 1&68 & 2&35 & 3&07 & 3&86 & 1&68 & 0&0 & 0&80 & 8&20  \\

$^{104}$Pd   & 1&36(27) & \omit\span & \omit\span & \omit\span &
0&61(8) &  33&3(74)      & \omit\span & \omit\span \\
           & 1&67 & 2&39 & 3&30 & 4&60 & 1&67 & 0&0 & 0&73 & 6&18 \\

$^{106}$Pd   & 1&63(28) & \omit\span & \omit\span & \omit\span &
0&98(12) &  26&2(31)      & 0&67(18) & \omit\span \\
           & 1&70 & 2&47 & 3&51 & 5&03 & 1&70 & 0&0 & 0&79 & 4&84 \\

$^{108}$Pd   & 1&47(20) & 2&16(28) & 2&99(48) & \omit\span &
1&43(14) &  16&6(18)      & 1&05(13) & 1&90(29) \\
           & 1&68 & 2&32 & 2&95 & 3&53 & 1&68 & 0&0 & 0&85 & 9&45 \\

$^{108}$Cd   & 1&54(24) & \omit\span & \omit\span & \omit\span &
0&64(20) &  67&7(120)      & \omit\span & \omit\span \\
           & 1&69 & 2&46 & 3&48 & 5&00 & 1&69 & 0&0 & 0&77 & 4&99 \\

$^{110}$Cd   & 1&68(24) & \omit\span & \omit\span & \omit\span &
1&09(19) &  48&9(78)      & \omit\span & 9&85(595) \\
           & 1&74 & 2&62 & 3&90 & 6&05 & 1&74 & 0&0 & 0&84 & 2&76 \\

$^{112}$Cd   & 2&02(22) & \omit\span & \omit\span & \omit\span &
0&50(10) &  19&9(35)      & 1&69(48) & 11&26(210) \\
           & 1&76 & 2&70 & 4&20 & 4&97 & 1&76 & 0&0 & 0&81 & 1&56 \\

$^{114}$Cd   & 1&99(25) & 3&83(72) & 2&73(97) & \omit\span &
0&71(24) &  15&4(29)      & 0&88(11) & 10&61(193) \\
           & 1&78 & 2&79 & 4&42 & 3&44 & 1&78 & 0&0 & 0&85 & 0&99 \\

\hline
\end{tabular}
\end{table}

\begin{table}

\setcounter{table}{1} \caption{ (continued) }

\bigskip

\begin{tabular}{l r@{.}l r@{.}l r@{.}l r@{.}l r@{.}l r@{.}l r@{.}l r@{.}l r@{.}l r@{.}l}

\hline
   \multicolumn{1}{l}{nucl.}
   &\multicolumn{2}{c} {$4_1\to 2_1 \over 2_1\to 0_1$}
    &\multicolumn{2}{c} {$6_1\to 4_1 \over 2_1\to 0_1$}
    &\multicolumn{2}{c} {$8_1\to 6_1 \over 2_1\to 0_1$}
   &\multicolumn{2}{c} {$10_1\to 8_1 \over 2_1\to 0_1$}
    &\multicolumn{2}{c} {$2_2 \to 2_1 \over 2_1\to 0_1$}
   &\multicolumn{2}{c}{$2_2 \to 0_1 \over 2_1\to 0_1$}
   &\multicolumn{2}{c}{$0_2 \to 2_1 \over 2_1\to 0_1$}
   &\multicolumn{2}{c}{$2_3 \to 0_1 \over 2_1 \to 0_1$}  \\

   & \omit\span & \omit\span & \omit\span & \omit\span &
  \omit\span &  \multicolumn{2}{c} {x $10^3$} &  \omit\span &
  \multicolumn{2}{c} {x $10^3$}
   \\
\hline

$^{116}$Cd   & 1&70(52) & \omit\span & \omit\span & \omit\span &
0&63(46) &  32&8(86)      & 0&02 & \omit\span \\
           & 1&66 & 2&39 & 3&43 & 5&00 & 1&66 & 0&0 & 0&63 & 4&55 \\

$^{118}$Cd   & $>$1&85 & \omit\span & \omit\span & \omit\span &
\omit\span & \omit\span      & 0&16(4) & \omit\span \\
           & 1&66 & 2&38 & 3&42 & 4&98 & 1&66 & 0&0 & 0&62 & 4&60 \\

$^{118}$Xe   & 1&11(7) & 0&88(27) & 0&49(20) & $>$0&73 &
\omit\span & \omit\span & \omit\span &\omit\span \\
           & 1&65 & 2&19 & 2&65 & 3&03 & 1&65 & 0&0 & 0&78 & 13&52 \\

$^{120}$Xe   & 1&16(14) & 1&17(24) & 0&96(22) & 0&91(19) &
\omit\span & \omit\span & \omit\span &\omit\span \\
           & 1&58 & 2&05 & 2&49 & 2&90 & 1&58 & 0&0 & 0&54 & 15&06 \\

$^{124}$Xe   & 1&34(24) & 1&59(71) & 0&63(29) & 0&29(8) &
0&70(19) &  15&9(46)      & \omit\span  & \omit\span \\
           & 1&57 & 2&05 & 2&50 & 2&92 & 1&57 & 0&0 & 0&54 & 14&87 \\

$^{128}$Xe   & 1&47(20) & 1&94(26) & 2&39(40) & 2&74(114) &
1&19(19) &  15&9(23)      & \omit\span  & \omit\span \\
           & 1&60 & 2&19 & 2&90 & 3&89 & 1&60 & 0&0 & 0&55 & 8&95 \\

$^{130}$Ba   & 1&36(6) & 1&62(15) & 1&55(56) & 0&93(15) &
\omit\span  & \omit\span   & \omit\span  & \omit\span \\
           & 1&66 & 2&28 & 2&90 & 3&50 & 1&66 & 0&0 & 0&78 & 10&07 \\

$^{132}$Ba   & \omit\span & \omit\span & \omit\span & \omit\span &
3&35(64) &  90&7(177)      & \omit\span & \omit\span \\
           & 1&48 &1&84 & 2&19 & 2&66 & 1&48 & 0&0 & 0&00 & 0&00 \\

$^{134}$Ba   & 1&55(21) & \omit\span & \omit\span & \omit\span &
2&17(69) &  12&5(41)      & \omit\span & \omit\span \\
           & 1&70 & 2&53 & 3&81 & 1&96 & 1&70 & 0&0 & 0&68 & 2&48 \\

$^{142}$Ba   & 1&40(17) & 0&56(14) & \omit\span & \omit\span &
\omit\span & \omit\span   & \omit\span & \omit\span \\
           & 1&50 & 1&89 & 2&30 & 2&88 & 1&50 & 0&0 & 0&11 & 2&37  \\

$^{148}$Nd   & 1&61(13) & 1&76(19) & \omit\span & \omit\span &
0&25(4) & 9&3(17)   & 0&54(6) & 32&82(816) \\
           & 1&63 & 2&14 & 2&56 & 2&89 & 1&63 & 0&0 & 0&73 & 15&09 \\

$^{152}$Gd   & 1&84(29) & 2&74(81) & \omit\span & \omit\span &
0&23(4) & 4&2(8)   & 2&47(78) & \omit\span \\
           & 1&71 & 2&47 & 3&39 & 4&54 & 1&71 & 0&0 & 0&85 & 5&63 \\

$^{154}$Dy   & 1&62(35) & 2&05(42) & 2&27(62) & 1&86(69) &
\omit\span   & \omit\span & \omit\span & \omit\span \\
           & 1&67 & 2&37 & 3&20 & 4&25 & 1&67 & 0&0 & 0&76 & 7&08 \\
\hline
\end{tabular}
\end{table}

\begin{table}

\setcounter{table}{1} \caption{ (continued) }

\bigskip

\begin{tabular}{l r@{.}l r@{.}l r@{.}l r@{.}l r@{.}l r@{.}l r@{.}l r@{.}l r@{.}l r@{.}l}

\hline
   \multicolumn{1}{l}{nucl.}
   &\multicolumn{2}{c} {$4_1\to 2_1 \over 2_1\to 0_1$}
    &\multicolumn{2}{c} {$6_1\to 4_1 \over 2_1\to 0_1$}
    &\multicolumn{2}{c} {$8_1\to 6_1 \over 2_1\to 0_1$}
   &\multicolumn{2}{c} {$10_1\to 8_1 \over 2_1\to 0_1$}
    &\multicolumn{2}{c} {$2_2 \to 2_1 \over 2_1\to 0_1$}
   &\multicolumn{2}{c}{$2_2 \to 0_1 \over 2_1\to 0_1$}
   &\multicolumn{2}{c}{$0_2 \to 2_1 \over 2_1\to 0_1$}
   &\multicolumn{2}{c}{$2_3 \to 0_1 \over 2_1 \to 0_1$}  \\

   & \omit\span & \omit\span & \omit\span & \omit\span &
  \omit\span &  \multicolumn{2}{c} {x $10^3$} &  \omit\span &
  \multicolumn{2}{c} {x $10^3$}
   \\
\hline

$^{156}$Er   & 1&78(16) & 1&89(36) & 0&76(20) & 0&88(22) &
\omit\span   & \omit\span & \omit\span & \omit\span \\
           & 1&62 & 2&23 & 2&91 & 3&74 & 1&62 & 0&0 & 0&64 & 9&60 \\

$^{192}$Pt   & 1&56(12) & 1&23(55) & \omit\span & \omit\span &
1&91(16)   & 9&5(9) & \omit\span & \omit\span \\
           & 1&58 & 2&08 & 2&55 & 3&00 & 1&58 & 0&0 & 0&56 & 14&25 \\

$^{194}$Pt   & 1&73(13) & 1&36(45) & 1&02(30) & 0&69(19) &
1&81(25) &  5&9(9)      & 0&01  & \omit\span \\
           & 1&58 & 2&06 & 2&49 & 2&88 & 1&58 & 0&0 & 0&58 & 15&34 \\

$^{196}$Pt   & 1&48(3) & 1&80(23) & 1&92(23) & \omit\span &
\omit\span &  0&4      & 0&07(4)  & 0&06(6) \\
           & 1&63 & 2&14 & 2&57 & 2&93 & 1&63 & 0&0 & 0&72 & 14&74 \\

$^{198}$Pt   & 1&19(13) & $>$1&78 & \omit\span & \omit\span &
 1&16(23) &  1&2(4)      & 0&81(22)  & 1&56(126) \\
           & 1&72 & 2&48 & 3&35 & 4&36 & 1&72 & 0&0 & 0&89 & 5&78 \\
\hline
\end{tabular}
\end{table}

\begin{table}

\caption{Comparison of experimental data \cite{NDS} (upper line) for several $B(E2)$ ratios of axially symmetric prolate deformed nuclei
to predictions (lower line) by the Bohr Hamiltonian with the Morse potential, for the
parameter values shown in Table \ref{MorsePar}.}\label{rotationalbe2}

\bigskip

\begin{tabular}{l r@{.}l r@{.}l r@{.}l r@{.}l r@{.}l r@{.}l r@{.}l r@{.}l r@{.}l r@{.}l}

\hline
   \multicolumn{1}{l}{nucl.}
   &\multicolumn{2}{c} {$4_1\to 2_1 \over 2_1\to 0_1$}
    &\multicolumn{2}{c} {$6_1\to 4_1 \over 2_1\to 0_1$}
    &\multicolumn{2}{c} {$8_1\to 6_1 \over 2_1\to 0_1$}
   &\multicolumn{2}{c} {$10_1\to 8_1 \over 2_1\to 0_1$}
    &\multicolumn{2}{c} {$2_\beta \to 0_1 \over 2_1\to 0_1$}
   &\multicolumn{2}{c}{$2_\beta \to 2_1 \over 2_1\to 0_1$}
   &\multicolumn{2}{c}{$2_\beta \to 4_1 \over 2_1\to 0_1$}
   &\multicolumn{2}{c}{$2_\gamma\to 0_1 \over 2_1 \to 0_1$}
   &\multicolumn{2}{c}{$2_\gamma\to 2_1 \over 2_1 \to 0_1$}
   &\multicolumn{2}{c}{$2_\gamma\to 4_1 \over 2_1 \to 0_1$} \\

   & \omit\span & \omit\span & \omit\span & \omit\span &
  \multicolumn{2}{c} {x $10^3$} &  \multicolumn{2}{c} {x $10^3$} &  \multicolumn{2}{c} {x $10^3$} &
  \multicolumn{2}{c} {x $10^3$} &  \multicolumn{2}{c} {x $10^3$} &  \multicolumn{2}{c} {x $10^3$}
   \\

\hline $^{154}$Sm   & 1&40(5) & 1&67(7) & 1&83(11) & 1&81(11) &
5&4(13) &  \omit\span      & \multicolumn{2}{c} {25(6)} &
18&4(34) &  \omit\span & 3&9(7) \\
           & 1&46 & 1&67 & 1&85 & 2&03 & 21&1 & 39&8 &  \multicolumn{2}{c}  {122} &47&3 & 69&2 & 3&6 \\

$^{156}$Gd & 1&41(5) & 1&58(6) & 1&71(10) & 1&68(9) &  3&4(3) &
\multicolumn{2}{c} {18(2)} & \multicolumn{2}{c} {22(2)} &
25&0(15) & 38&7(24) & 4&1(3)  \\
           & 1&47 & 1&70 & 1&90 & 2&12 & 22&5 & 44&3 & \multicolumn{2}{c} {145} &63&0 & 92&4  & 4&9 \\

$^{158}$Gd & 1&46(5) &  \omit\span        & 1&67(16) & 1&72(16) &
1&6(2) & 0&4(1) & 7&0(8) &
17&2(20) & 30&3(45) & 1&4(2) \\
           & 1&45 & 1&64 & 1&79 & 1&95 & 14&6 & 26&7 & \multicolumn{2}{c} {79}  &63&3 & 91&7 &  4&7 \\

$^{158}$Dy & 1&45(10) & 1&86(12) & 1&86(38) & 1&75(28) &
\multicolumn{2}{c} {12(3)} & \multicolumn{2}{l} {19(4)} &
\multicolumn{2}{c} {66(16)} &
32&2(78) & 103&8(258) & 11&5(48) \\
           & 1&48 & 1&73 & 1&98 & 2&26 & 23&5 & 49&0 & \multicolumn{2}{c} {175} &95&1 & 140&1 & 7&5 \\

$^{160}$Dy & 1&46(7) & 1&23(7) & 1&70(16) & 1&69(9) &  3&4(4) &
\omit\span   & 8&5(10) &
23&2(21) & 43&8(42) & 3&1(3) \\
           & 1&46 & 1&67 & 1&84 & 2&02 & 20&9 & 39&3 & \multicolumn{2}{c} {120} &82&7 & 120&1 & 6&2 \\

$^{162}$Dy & 1&45(7) & 1&51(10) & 1&74(10) & 1&76(13) & \omit\span
&  \omit\span   &  \omit\span   &
0&12(1) & 0&20  & 0&02\\
           & 1&45 & 1&62 & 1&76 & 1&88 & 11&4 & 20&1 & \multicolumn{2}{c} {56}  &94&3 & 135&6 & 6&9 \\

$^{164}$Dy & 1&30(7) & 1&56(7) & 1&48(9) & 1&69(9) &  \omit\span &
\omit\span &  \omit\span   &
19&1(22) & 38&3(39) & 4&6(5) \\
           & 1&44 & 1&62 & 1&74 & 1&85 & 16&1 & 27&6 & \multicolumn{2}{c} {72} &99&7 & 143&2 & 7&3 \\

$^{162}$Er &  \omit\span         &  \omit\span                &
\omit\span  & \omit\span     & \multicolumn{2}{c} {8(7)} &
\omit\span  & \multicolumn{2}{c} {170(90)} &
32&5(28) & 77&0(56) &  9&4(69) \\
           & 1&47 & 1&72 & 1&95 & 2&21 & 22&0 & 45&1 & \multicolumn{2}{c} {158} &96&6 & 141&7 & 7&5 \\

$^{164}$Er & 1&18(13) &   \omit\span        & 1&57(9) & 1&64(11) &
\omit\span &  \omit\span  &  \omit\span   &
23&9(35) & 52&3(72) & 7&8(12) \\
           & 1&47 & 1&69 & 1&87 & 2&05 & 23&4 & 44&8 & \multicolumn{2}{c} {139} &103&3 & 150&3 & 7&8 \\

$^{166}$Er & 1&45(12) & 1&62(22) & 1&71(25) & 1&73(23) &
\omit\span    &  \omit\span   &  \omit\span   &
25&7(31) & 45&3(54) & 3&1(4) \\
           & 1&45 & 1&64 & 1&78 & 1&92 & 17&3 & 30&9 & \multicolumn{2}{c} {88}  &104&5 & 150&7 & 7&7 \\

$^{168}$Er & 1&54(7) & 2&13(16) & 1&69(11) & 1&46(11) & \omit\span
&  \omit\span   &  \omit\span   &
23&2(15) & 41&1(31) & 3&0(3) \\
           & 1&44 & 1&61 & 1&73 & 1&84 & 1&8 & 3&1 & \multicolumn{2}{c} {8} & 101&1 & 145&1 & 7&3 \\

$^{170}$Er &  \omit\span         &  \omit\span         & 1&78(15)
& 1&54(11) & 1&4(1) & 0&2(2) & 6&8(12) &
17&7(9) &  \omit\span        & 1&4(4) \\
           & 1&46 & 1&68 & 1&84 & 1&98 & 28&1 & 50&6 & \multicolumn{2}{c} {139} &76&1 & 110&6 & 5&7 \\

\hline
\end{tabular}
\end{table}

\begin{table}
\setcounter{table}{2} \caption{ (continued) }

\bigskip

\begin{tabular}{l r@{.}l r@{.}l r@{.}l r@{.}l r@{.}l r@{.}l r@{.}l r@{.}l r@{.}l r@{.}l}

\hline
   \multicolumn{1}{l}{nucl.}
   &\multicolumn{2}{c} {$4_1\to 2_1 \over 2_1\to 0_1$}
    &\multicolumn{2}{c} {$6_1\to 4_1 \over 2_1\to 0_1$}
    &\multicolumn{2}{c} {$8_1\to 6_1 \over 2_1\to 0_1$}
   &\multicolumn{2}{c} {$10_1\to 8_1 \over 2_1\to 0_1$}
    &\multicolumn{2}{c} {$2_\beta \to 0_1 \over 2_1\to 0_1$}
   &\multicolumn{2}{c}{$2_\beta \to 2_1 \over 2_1\to 0_1$}
   &\multicolumn{2}{c}{$2_\beta \to 4_1 \over 2_1\to 0_1$}
   &\multicolumn{2}{c}{$2_\gamma\to 0_1 \over 2_1 \to 0_1$}
   &\multicolumn{2}{c}{$2_\gamma\to 2_1 \over 2_1 \to 0_1$}
   &\multicolumn{2}{c}{$2_\gamma\to 4_1 \over 2_1 \to 0_1$} \\

   & \omit\span & \omit\span & \omit\span & \omit\span &
  \multicolumn{2}{c} {x $10^3$} &  \multicolumn{2}{c} {x $10^3$} &  \multicolumn{2}{c} {x $10^3$}
   & \multicolumn{2}{c} {x $10^3$} &  \multicolumn{2}{c} {x $10^3$} &  \multicolumn{2}{c} {x $10^3$}\\
\hline

$^{166}$Yb & 1&43(9) & 1&53(10) & 1&70(18) & 1&61(80) & \omit\span
&  \omit\span   &  \omit\span   &
     \omit\span    &  \omit\span    &  \omit\span        \\
           & 1&48 & 1&73 & 1&97 & 2&23 &  24&3 & 50&3 & \multicolumn{2}{c} {176} &101&4 & 149&0 & 7&9 \\

$^{168}$Yb &  \omit\span         &   \omit\span             &
\omit\span  & \omit\span       & 8&6(9) &  \omit\span   &
\omit\span   &
22&0(55) & 45&9(73) & 8&6 \\
           & 1&47 & 1&70 & 1&90 & 2&11 & 23&3 & 45&7 & \multicolumn{2}{c} {148} & 84&6 & 123&7 & 6&5  \\

$^{170}$Yb &  \omit\span         &  \omit\span        & 1&79(16) &
1&77(14) & 5&4(10) &  \omit\span   &  \omit\span   &
13&4(34) & 23&9(57) & 2&4(6) \\
           & 1&44 & 1&64 & 1&78 & 1&93 & 8&3  & 15&1 & \multicolumn{2}{c} {44}  &73&6 & 106&3 & 5&4 \\

$^{172}$Yb & 1&42(10) & 1&51(14) & 1&89(19) & 1&77(11) & 1&1(1) &
3&7(6) & \multicolumn{2}{c} {12(1)} & 6&3(6) & \omit\span   & 0&6(1) \\
           & 1&46 & 1&67 & 1&83 & 1&99 & 23&6 & 43&1 & \multicolumn{2}{c} {124} &48&7 & 70&9 & 3&7 \\

$^{174}$Yb & 1&39(7) & 1&84(26) & 1&93(12) & 1&67(12) & \omit\span
&  \omit\span  &  \omit\span  &
    \omit\span  & 12&4(29)   & \omit\span          \\
           & 1&45 & 1&63 & 1&75 & 1&86 & 18&9 & 32&3 & \multicolumn{2}{c} {88} &44&6 & 64&3 & 3&3 \\

$^{176}$Yb & 1&49(15) & 1&63(14) & 1&65(28) & 1&76(18) &
\omit\span    &  \omit\span   &  \omit\span   &9&8
 &  \omit\span   &  \omit\span       \\
           & 1&44 & 1&62 & 1&74 & 1&87 & 0&5  & 1&0  & \multicolumn{2}{c} {3} & 65&1 & 93&7  & 4&8 \\

$^{174}$Hf &  \omit\span      &   \omit\span       &
\omit\span &  \omit\span  & \multicolumn{2}{c}{14(4)} & \omit\span& \multicolumn{2}{c} {9(3)} &31&6(161) & 48&7(124) & \omit\span        \\
           & 1&46 & 1&66 & 1&81 & 1&97 & 20&3 & 37&3 & \multicolumn{2}{c} {109} & 65&4 & 94&9 & 4&9 \\

$^{176}$Hf & \omit\span  & \omit\span      & \omit\span       &
\omit\span & 5&4(11) & \omit\span & \multicolumn{2}{c} {31(6)} &21&3(26) & \omit\span  &   \omit\span     \\
           & 1&46 & 1&68 & 1&86 & 2&04 & 23&3 & 43&9 & \multicolumn{2}{c} {133} &57&0 & 83&2 & 4&4 \\

$^{178}$Hf & \omit\span       & 1&38(9) & 1&49(6) & 1&62(7) &
0&4(2) & \omit\span & 2&4(9) &24&5(39) & 27&7(28) & 1&6(2) \\
           & 1&46 & 1&68 & 1&86 & 2&04 & 23&3 & 44&3 & \multicolumn{2}{c} {136} &73&5 & 107&2 & 5&6 \\

$^{182}$W  & 1&43(8) & 1&46(16) & 1&53(14) & 1&48(14) &  6&6(6) &
4&6(6) & \multicolumn{2}{c} {13(1)} &24&8(12) & 49&2(24) & 0&2\\
           & 1&46 & 1&68 & 1&85 & 2&00 & 25&7 & 47&4 & \multicolumn{2}{c} {137} &79&3 & 115&4 & 6&0 \\

$^{184}$W  & 1&35(12) & 1&54(9) & 2&00(18) & 2&45(51) & 1&8(3) &
\omit\span & \multicolumn{2}{c} {24(3)} &37&1(28) & 70&6(51) & 4&0(4) \\
           & 1&48 & 1&71 & 1&88 & 2&04 & 30&5 & 57&4 & \multicolumn{2}{c} {167} & 124&8 & 182&0 & 9&4 \\

$^{186}$W  & 1&30(9) & 1&69(12) & 1&60(12) & 1&36(36) & \omit\span & \omit\span
& \omit\span & 41&7(92) & 91&0(201) & \omit\span        \\
           & 1&49 & 1&76 & 1&99 & 2&20 & 31&4 & 64&7 & \multicolumn{2}{c} {213} &164&0 & 241&2 & 12&7 \\

$^{186}$Os & 1&45(7) & 1&99(7) & 1&89(11) & 2&06(44) & \omit\span
& \omit\span & \omit\span &109&4(71) & 254&6(150) & 13&0(47) \\
           & 1&50 & 1&80 & 2&11 & 2&45 & 26&2 & 59&5 & \multicolumn{2}{c} {235} &163&5 & 242&2 & 13&0 \\

\hline
\end{tabular}
\end{table}

\begin{table}
\setcounter{table}{2} \caption{ (continued) }

\bigskip

\begin{tabular}{l r@{.}l r@{.}l r@{.}l r@{.}l r@{.}l r@{.}l r@{.}l r@{.}l r@{.}l r@{.}l}

\hline
   \multicolumn{1}{l}{nucl.}
   &\multicolumn{2}{c} {$4_1\to 2_1 \over 2_1\to 0_1$}
    &\multicolumn{2}{c} {$6_1\to 4_1 \over 2_1\to 0_1$}
    &\multicolumn{2}{c} {$8_1\to 6_1 \over 2_1\to 0_1$}
   &\multicolumn{2}{c} {$10_1\to 8_1 \over 2_1\to 0_1$}
    &\multicolumn{2}{c} {$2_\beta \to 0_1 \over 2_1\to 0_1$}
   &\multicolumn{2}{c}{$2_\beta \to 2_1 \over 2_1\to 0_1$}
   &\multicolumn{2}{c}{$2_\beta \to 4_1 \over 2_1\to 0_1$}
   &\multicolumn{2}{c}{$2_\gamma\to 0_1 \over 2_1 \to 0_1$}
   &\multicolumn{2}{c}{$2_\gamma\to 2_1 \over 2_1 \to 0_1$}
   &\multicolumn{2}{c}{$2_\gamma\to 4_1 \over 2_1 \to 0_1$} \\

   & \omit\span & \omit\span & \omit\span & \omit\span &
  \multicolumn{2}{c} {x $10^3$} &  \multicolumn{2}{c} {x $10^3$} &  \multicolumn{2}{c} {x $10^3$}
   & \multicolumn{2}{c} {x $10^3$} &  \multicolumn{2}{c} {x $10^3$} &  \multicolumn{2}{c} {x $10^3$}   \\
\hline

$^{188}$Os & 1&68(11) & 1&75(11) & 2&04(15) & 2&38(32)        &
\omit\span & \omit\span & \omit\span &63&3(92) & 202&5(304) & 43&0(74) \\
           & 1&51 & 1&84 & 2&22 & 2&73 & 23&0 & 56&7 & \multicolumn{2}{c} {257} &245&3 & 363&5 & 19&4 \\

$^{230}$Th & 1&36(8)         & \omit\span      & \omit\span &
\omit\span & 5&7(26) & \omit\span & \multicolumn{2}{c} {20(11)} &15&6(59) & 28&1(100) & 1&8(11) \\
           & 1&47 & 1&69 & 1&88 & 2&07 & 23&6 & 45&3 & \multicolumn{2}{c} {141} & 62&5 & 91&4 & 4&8 \\

$^{232}$Th & 1&44(15) & 1&65(14) & 1&73(12) & 1&82(15) &
\multicolumn{2}{c}{14(6)} & 2&6(13) & \multicolumn{2}{c} {17(8)} &14&6(28) & 36&4(56) & 0&7 \\
           & 1&46 & 1&66 & 1&82 & 1&98 & 21&3 & 39&1 & \multicolumn{2}{c} {115} & 56&7 & 82&4 & 4&3 \\

$^{234}$U  & \omit\span &   \omit\span     & \omit\span
&\omit\span &\omit\span &\omit\span & \omit\span &12&5(27) & 21&1(44) & 1&2(3) \\
           & 1&45 & 1&63 & 1&77 & 1&89 & 18&5 & 32&4 & \multicolumn{2}{c} {88} & 42&8 & 62&0 & 3&2 \\

$^{236}$U  & 1&42(11) & 1&55(11) & 1&59(17) & 1&46(17) &
\omit\span & \omit\span & \omit\span & \omit\span      &   \omit\span    &   \omit\span    \\
           & 1&45 & 1&63 & 1&75 & 1&87 & 17&6 & 30&4 & \multicolumn{2}{c} {80} & 44&6 & 64&4 & 3&3 \\

$^{238}$U  & \omit\span       & \omit\span       & 1&45(23) &
1&71(22) & 1&4(6) & 3&6(14) & \multicolumn{2}{c}{12(5)} &10&8(8) & 18&9(17)  & 1&2(1) \\
           & 1&45 & 1&62 & 1&75 & 1&86 & 17&2 & 29&6 & \multicolumn{2}{c} {77} & 37&7 & 54&4 & 2&8 \\

$^{238}$Pu &  \omit\span            &  \omit\span     & \omit\span
& \omit\span& \multicolumn{2}{c}{14(4)} &\omit\span &
\multicolumn{2}{c}{11(4)} &   \omit\span   &  \omit\span   &  \omit\span      \\
           & 1&44 & 1&60 & 1&70 & 1&78 & 6&1 & 10&0 & \multicolumn{2}{c} {24} & 42&4 & 60&8 & 3&1 \\

$^{250}$Cf &  \omit\span            &  \omit\span     & \omit\span
&\omit\span & \omit\span&\omit\span &\omit\span &6&8(17) & 10&9(25) & 0&6(1) \\
           & 1&44 & 1&60 & 1&71 & 1&80 & 13&0 & 21&6 & \multicolumn{2}{c} {53} & 40&1 & 57&7 & 2&9 \\
\hline
\end{tabular}
\end{table}

\end{document}